\documentclass[aps,prd,nofootinbib,showpacs,superscriptaddress, preprint]{revtex4-1}
\pdfoutput=1
\usepackage[utf8]{inputenc}
\usepackage{amsmath,amssymb}
\usepackage{epsfig}
\usepackage{graphicx}
\usepackage[usenames,dvipsnames]{color}
\usepackage[colorlinks,citecolor=blue]{hyperref}
\usepackage{soul}
\usepackage{color}
\usepackage{subfigure}

\begin{document}

\title{Soft leptogenesis in the NMSSM with a singlet right-handed neutrino superfield}

\author{Waleed Abdallah}
\email{waleedabdallah@hri.res.in}
\affiliation{Harish-Chandra Research Institute, Chhatnag Road, Jhunsi, Allahabad~ 211 019, India}
\affiliation{Department of Mathematics, Faculty of Science, Cairo University, Giza 12613, Egypt}
\author{Abhass Kumar}
\email{abhass@prl.res.in}
\affiliation{Theoretical Physics Division, Physical Research Laboratory, Navrangpura, Ahmedabad~380 009, India}
\author{Abhijit Kumar Saha}
\email{aks@prl.res.in}
\affiliation{Theoretical Physics Division, Physical Research Laboratory, Navrangpura, Ahmedabad~380 009, India}

\begin{abstract}
In this work, we explore soft leptogenesis in the NMSSM framework extended by a right-handed neutrino superfield. We calculate the CP asymmetry, $\varepsilon$, and find it to be non-zero at tree{-}level without using thermal effects for the final state particles. This is in contrast to soft leptogenesis in the MSSM extended by a right-handed neutrino superfield where thermal effects are essential. The difference arises due to the presence of a 3-body decay of the sneutrino in the NMSSM that violates lepton number at tree{-}level. Apart from this, we also find that $\varepsilon\neq 0$ if the additional singlet scalar has a complex vacuum expectation value while all the other NMSSM parameters including the soft SUSY breaking ones relevant for CP asymmetry remain real. We estimate the order of magnitudes of these parameters to produce sufficient baryon asymmetry of the Universe.
\end{abstract}
\maketitle

\section{Introduction}
It is well known that the observable Universe has an asymmetry between baryons and anti-baryons~\cite{Tanabashi:2018oca}, often called the problem of baryogenesis. Over the years, many mechanisms have been proposed to create this baryon asymmetry. More recently, leptogenesis~\cite{Fukugita:1986hr,Strumia:2006qk} has become a highly favoured model for baryogenesis, specially because this mechanism is naturally linked to neutrino masses. Adding right-handed (RH) singlet heavy neutrinos to the standard model {(SM)} generates neutrino masses by the seesaw mechanism~\cite{Minkowski:1977sc, Mohapatra:1979ia, Yanagida:1979as, GellMann:1980vs, Glashow:1979nm, Schechter:1980gr}. These RH neutrinos can also decay to produce a scalar and SM leptons and if the decay violates CP for example due to interference between tree-level and loop-level decays owing to complex couplings, a lepton asymmetry is generated. Since, in the {SM}, the $B-L$ symmetry is exact while the $B+L$ symmetry is broken by the electroweak (EW) sphaleron processes~\citep{Kuzmin:1985mm}, these sphaleron processes can convert the generated lepton asymmetry to baryon asymmetry.

Soft leptogenesis~\cite{Grossman:2003jv, DAmbrosio:2003nfv,Fong:2011yx} pertains to generating lepton asymmetry at the tree-level itself due to mixing between the particle and anti-particle states of the RH singlet sneutrino, $\tilde{N}$, because of the presence of soft SUSY breaking terms\footnote{Soft leptogenesis in different types of SUSY framework can be found in~\cite{Chen:2004xy,Chun:2005ms,Medina:2006hi,Garayoa:2006xs,Fong:2008mu,Fong:2009iu,Kajiyama:2009ae,Fong:2010zu,Hamaguchi:2010cw,Fong:2010bv,Kuismanen:2012iz}.}. In the most minimal soft leptogenesis setup using minimal supersymmetric standard model (MSSM)~\cite{Chung:2003fi,Martin:1997ns} extended by one RH neutrino superfield, $\hat{N}$, a CP asymmetry in the {RH} sneutrino sector is created only when  thermal masses for the final products {are considered}. The asymmetry is present because the thermal phase space factors are different for bosons and fermions. Another work featuring soft leptogenesis looks at CP violation not just due to mixing between particle and anti-particle initial states but in decays and in the interference of mixing and decay~\cite{Grossman:2004dz}. In {Ref.}~\cite{Adhikari:2015ysa}, it is shown that considering most generic soft trilinear couplings and one loop self energy contributions for sneutrino decay it is possible to generate CP violation even without finite temperature effects within the same {setup}.
 
However, the MSSM suffers from the so-called $\mu$-problem~\cite{Kim:1983dt} -- there is no explanation to why the SUSY scale preserving $\mu$-term (a direct SUSY mass term for the Higgs fields) should be of the same order as the soft SUSY breaking terms. The most straightforward solution to the $\mu$-problem comes by promoting the $\mu$-parameter into a field whose vacuum expectation value (vev) is determined, like the other scalar field {vevs}, from the minimization of the scalar potential along the new field direction~\cite{Ellwanger:2009dp}. Naturally, it is expected to fall in the range of the other vevs, i.e., of order ${\cal O}(M_{\rm SUSY})$. The next-to-minimal supersymmetric standard model (NMSSM) (for review see~\cite{Ellwanger:2009dp,Maniatis:2009re}) is the most simple and elegant model to solve this problem, where a singlet superfield $\hat S$ is introduced to the MSSM superfields which gets non{-}zero vev. The NMSSM can be extended by a set of {RH} neutrino superfields to generate masses for the SM light neutrinos by the type-I seesaw (see~\cite{Grossman:1997is,King:1998jw,Kitano:1999qb} for {the} MSSM extended by RH neutrino superfield). This has been explored earlier in {Ref.}~\cite{Das:2010wp}. This extension also keeps the R-parity conserved if the  sneutrinos do not get vevs~\cite{Das:2010wp}. 

In this work, by using the NMSSM extended by the {RH} neutrino superfield, we present a soft leptogenesis scenario that creates a lepton asymmetry at the tree-level decay of the {RH} sneutrino without using thermal mass factors. The CP violation is achieved by the mixing between the particle and anti-particle states. This is due to the presence of the soft terms and the trilinear coupling between the additional singlet superfield which takes a vev and the RH neutrino superfield. A similar non-SUSY setup with such a trilinear term can be found in~\cite{Alanne:2018brf}. We also show that it is possible to obtain non{-}zero CP asymmetry even when all the soft parameters are real. Since the soft terms are responsible for creating the CP asymmetry instead of needing flavour effects as in usual leptogenesis, using only one generation of the RH neutrino superfield is enough. Even so, the {setup} can be easily extended to get the experimentally observed SM neutrino mass hierarchies and their mixing angles pattern~\cite{Esteban:2016qun,deSalas:2017kay}.

The paper is organised in the following manner. We setup the model and segregate the parts required for soft leptogenesis in the next section (Sec.~\ref{sec:model}). In the one following that{,} i.e.{,} Sec.~\ref{sec:CP}, we calculate the CP asymmetry produced by decays of the various particle present in the model that contribute to non-zero CP asymmetry parameter $\varepsilon$ at the tree{-}level. We talk about the decays of $\tilde{N}$ as well as the scalar $S$ in the model. In Sec.~\ref{sec:consts}, we discuss the most crucial and important constraints and give a simple expression for $\varepsilon$. In Sec.~\ref{sec:results}, we give and discuss the results of our calculation. We find that for successfully generating the observed baryon asymmetry of the Universe, we need $\varepsilon\approx \mathcal{O}(10^{-6})$. We also discuss what this could mean for various parameters of the model including the soft ones. We finally conclude in Sec.~\ref{sec:conclude}.

\section{Model}\label{sec:model}
In the NMSSM, an extra singlet superfield $\hat{S}$ is added to the MSSM
Higgs sector~\cite{Ellwanger:2009dp}. Assuming explicitly $\mathbb{Z}_3$ symmetry, the superpotential for the NMSSM with a singlet {RH} neutrino superfield $\hat{N}$ in terms of the new singlet superfield $\hat{S}$ and the MSSM doublet superfields $\hat{H}_u$ and $\hat{H}_d$ will be as follows~\cite{Ellwanger:2009dp}:
\begin{align}
W= Y^{ij}_E \hat H_d \hat L_i \hat E_j + Y_D^{ij} \hat H_d \hat Q_i \hat D_j + Y_U^{ij} \hat H_u \hat Q_i \hat U_j + \lambda \hat S \hat H_u \hat H_d +\frac{\kappa}{3} \hat S^3+ Y^i_N \hat{N} \hat H_u \hat L_i  + \lambda_N \hat S \hat N\hat N ,\label{eq:Sup}
\end{align}
where $\hat L_i$ and $\hat Q_i$ are the $SU(2)$ doublet superfields of leptons and quarks; $\hat E_i$ and $\hat D_i$ $(\hat U_i)$ denote singlet down (up)-type quark superfields, respectively, and $Y$'s, $\lambda$'s and $\kappa$ are dimensionless couplings with generation indices ($i,j=1,2,3$). 
After the singlet $S$ obtains a vacuum expectation value (vev) $\langle S\rangle$, an effective $\mu$-term is generated: $\mu_{\rm eff}=\lambda \langle S\rangle$, which solves the so-called $\mu$-problem~\cite{Kim:1983dt}. The soft SUSY-breaking Lagrangian is given by
\begin{align}
-{\cal L}_{\rm soft} &=-{\cal L}_{\rm soft}^{\rm MSSM}|_{B\mu=0}+ \left(A_{\lambda_H} \lambda S  H_u H_d +A_\kappa\frac{\kappa}{3} S^3+A_N^i Y^i_N \tilde N  H_u \tilde L_i  +  A_\lambda \lambda_N S \tilde N \tilde N  + {\rm h.c.}\right) \nonumber\\
&+m_S^2 |S|^2 + M^2 {|\tilde N|^2},
\end{align}
where $\tilde{L}_{i}$ and $\tilde{N}$ are the scalar components of $\hat L_{i}$ and $\hat N$ superfields, respectively. CP is spontaneously violated when the scalars $H_u,H_d,S$ attain vevs with relative physical phases. The vev of the singlet $S$ is complex:
\begin{equation}
\langle S\rangle = v_S e^{i\delta}{.}
\end{equation}
Since leptogenesis occurs above the electroweak (EW) phase transition, we do not give {vevs} to the two Higgs doublets. In this case, spontaneous CP violation can occur only when $\sin\delta \neq 0$.

\subsection{Terms relevant for soft leptogenesis}
The terms from the superpotential required for leptogenesis via sneutrino decay are:
\begin{align}
 W~{\supset}~Y_{N} \hat L \hat H \hat N+ \lambda_N {\hat S} \hat N \hat N +\frac{\kappa}{3} {\hat S^3}.
\end{align}
Here we consider $\lambda_N,~\kappa$ to be {all} real and positive. We also remove the $i,j$ indices from the leptons and the {$u$} index from the Higgs superfield for brevity.
The scalar potential is obtained using:
\begin{align}
 V_S=\Big|\frac{\partial W}{\partial S}\Big|^2+\Big|\frac{\partial W}{\partial N}\Big|^2+\Big|\frac{\partial W}{\partial L}\Big|^2,
\end{align}
with
\begin{align}
 \Big|\frac{\partial W}{\partial S}\Big|^2 &=\lambda_N^2|\tilde{N}|^4+\kappa^2 |S|^4+\lambda_N\; \kappa S^{*2} \tilde{N}\tilde{N}+\lambda_N\;\kappa \tilde{N}^*\tilde{N}^* S^2,\\
 \Big|\frac{\partial W}{\partial N}\Big|^2 &= |Y_{N}|^2|\tilde{L}|^2|H|^2+4 \lambda_N^2|\tilde{N}|^2|S|^2+2\lambda_N Y^*_{N} \tilde{N} S \tilde{L}^*H^*+2\lambda_N Y_{N} \tilde{N}^* S^* H \tilde{L},\\
 \Big|\frac{\partial W}{\partial L}\Big|^2 &=|Y_{N}|^2 |H|^2 |\tilde{N}|^2.
\end{align}
The fermionic part of the Lagrangian is given by:
\begin{align}
 \mathcal{L}_f= Y_{N}  L \tilde{H} \tilde{N}+\lambda_N S N N + Y_{N} L H N.
\end{align}
The soft {SUSY-}breaking {Lagrangian} terms that play a role in leptogenesis are:
\begin{align}
 {-{\cal L}_{\rm soft}}~{\supset}~\left( A_\kappa \frac{\kappa}{3} {S^3} +A_N Y_{N} \tilde{L} H \tilde{N}+A_\lambda \lambda_N S \tilde{N}\tilde{N}+ h.c.\right)+m_S^2 {|S|^2} + M^2 {|\tilde N|^2}. 
\end{align}
The superpotential and the soft breaking terms combine to give the following interactions for $\tilde{N}$ and $\sigma$ which could in principle contribute to soft leptogenesis due to mixing between the particle and anti-particle states through the soft terms:
\begin{align}\label{eq:interaction}
\mathcal{L}_{\rm int} &= \tilde{N}\left( Y_{N} \tilde{H} L + 2 \lambda_N Y^*_{N} v_S e^{i\delta} H^* \tilde{L}^* + 2 \lambda_N Y^*_{N} \sigma H^* \tilde{L}^* + A_N Y_{N} H \tilde{L}\right)\nonumber
\\ &+ \sigma \left(\lambda_N NN + A_\lambda \lambda_N \tilde{N}\tilde{N}\right) + h.c.,
\end{align}
where $\sigma = S- \langle S \rangle$.
\section{CP asymmetry}\label{sec:CP}
Because of the soft terms as well as the vev of $S$, there is a mixing between particle and anti-particle states of the sneutrino and the singlet scalar $\sigma$ which is the dynamic part of $S$. The {squared mass} matrices for the two of them are {given by}:
\begin{eqnarray}
{\cal M}_{\tilde{N}}^2 &=&
 \begin{bmatrix}
    {M_1^2} & \lambda_N \kappa v_S^2 e^{2i\delta}+A_\lambda \lambda_N v_S e^{-i\delta}  \\
    \lambda_N \kappa v_S^2 e^{-2i\delta}+A_\lambda \lambda_N v_S e^{i\delta} & {M_1^2} 
\end{bmatrix},\label{eq:M2N} \\ \nonumber\\
{\cal M}_\sigma ^2 &=& \begin{bmatrix}
{m_\sigma^2}  & 2  \kappa^2 v^2_{S} e^{2 i \delta}+2A_\kappa^* \kappa v_{S} e^{-i\delta} \\
2   \kappa^2 v^2_{S} e^{-2 i \delta}+2A_\kappa \kappa v_{S} e^{i\delta} & {m_\sigma^2}\label{eq:M2S}
\end{bmatrix}{,}
\end{eqnarray}
{where}
\begin{eqnarray}
M_1^2 &=& M^2+4 \lambda_N^2 v_S^2, \nonumber\\
m_\sigma^2 &=& m^2_{S}+4 \kappa^2 v^2_{S}.
\end{eqnarray}
If $A_\lambda$ is real, the mass square eigenvalues of the sneutrino are:
\begin{align}
M^2_\pm = {M^2_1} \pm \sqrt{A_\lambda^2 \lambda_N^2 v^2_{S} +\lambda_N^2 \kappa^2 v^4_{S} + 2 A_\lambda \lambda_N^2 v^3_{S} \kappa \cos(3\delta)}~,
\end{align}
which have the following eigenstates:
\begin{equation}
\tilde{N}_{\pm} = \frac{1}{\sqrt{2}}\left(\tilde{N} \pm\tilde{N}^*  \right){.}
\end{equation}
Similarly, one can write the mass square eigenvalues and eigenstates for the $\sigma-\sigma^*$ system. Because of mixing between the particle and anti-particle states of sneutrino and singlet scalar, these systems are similar to $K_0-\bar{K}_0$ and $B_0-\bar{B}_0$ systems~\cite{Nir:2001ge}. The evolution of these systems in the non-relativistic limit are driven by the Hamiltonian ${\cal H}$ defined as follows:
\begin{align}
{\cal H}={\cal M}-\frac{i}{2}\Gamma, \label{eq:Hamiltonian}
\end{align}
where ${\cal M}$ is the mass matrix and $\Gamma$ is the decay rate matrix of the corresponding system.

Finally, the decay rates of the time evolved particle and anti-particle states of the $\tilde{N}-\tilde{N}^*$ and the $\sigma-\sigma^*$ system are calculated to get the final total CP asymmetry. As can be seen from Eq.~(\ref{eq:interaction}), both $\tilde{N}$ and $\sigma$ can decay to produce lepton asymmetry depending on the nature of their couplings. Therefore, we separately consider the two limiting cases where only one of them can decay at a time by fixing their masses. The first case is when $M_1\gg m_\sigma$. In this case, $\tilde{N}$ decays to produce the CP asymmetry at the tree-level while decays of $\sigma$ into a pair of RH (s)neutrinos are kinematically suppressed. The relevant Feynman diagrams for this situation are shown in Figs.~\ref{fig:FD-SN} and \ref{fig:FD-SN1}. {Fig. \ref{fig:FD-SN} shows the point interaction, 2 and 3 body decays of $\tilde{N}$ while Fig. \ref{fig:FD-SN1} shows the other possible 3-body decays of $\tilde{N}$ that are mediated by an off-shell $\tilde{N}$. The first three point interactions of Fig. \ref{fig:FD-SN} are present in MSSM soft-leptogenesis as well and the observed matter-anti-matter asymmetry requires the decay products to have thermal corrections. The last diagram of Fig. \ref{fig:FD-SN} and those in Fig. \ref{fig:FD-SN1} are presnt only in NMSSM. We will see in Sec.~\ref{sec:Ndecays} that the 3-body decay diagram of Fig. \ref{fig:FD-SN} is responsible for creating enough CP asymmetry without any thermal mass corrections. The contribution of diagrams in Fig. \ref{fig:FD-SN1} is suppressed by the higher order soft terms.} The second possibility to create the CP asymmetry is through decays of $\sigma$ which occurs when $m_\sigma\gg M_1$. The diagrams are shown in Fig.~\ref{fig:FD-sigma}. However, as we will see in Sec.~\ref{sec:sigmadecay}, these decays can not produce any CP asymmetry at the tree-level via the same mechanism followed by $\tilde{N}$ decays. This tells us that we can ignore the decays of $\sigma$ without any loss of generality for the study of soft leptogenesis in the present setup. We consider these cases one-by-one.

\subsection{$\tilde{N}$ decays}\label{sec:Ndecays}
\begin{figure}[h]
\begin{center}
\includegraphics[height=3cm,width=4cm]{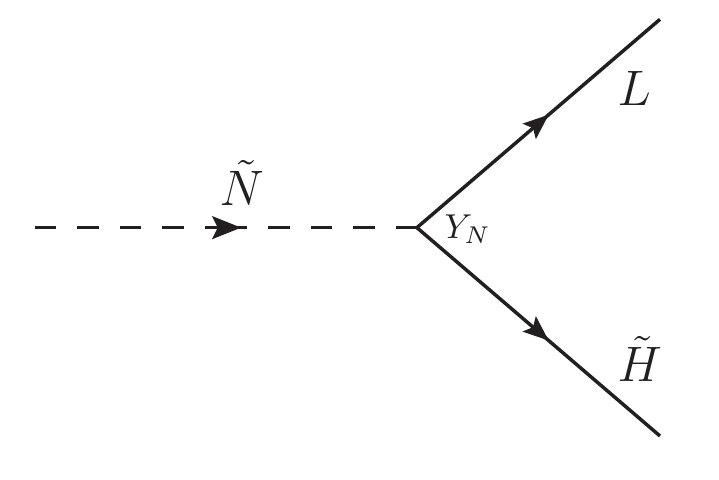}
\includegraphics[height=3cm,width=4cm]{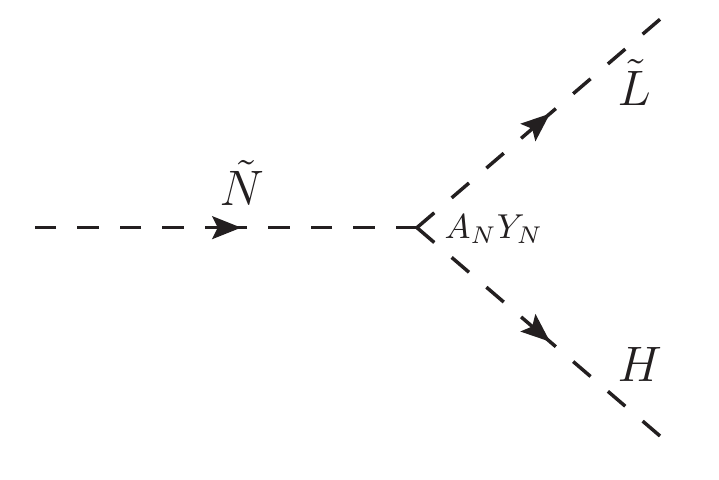}\includegraphics[height=3cm,width=4cm]{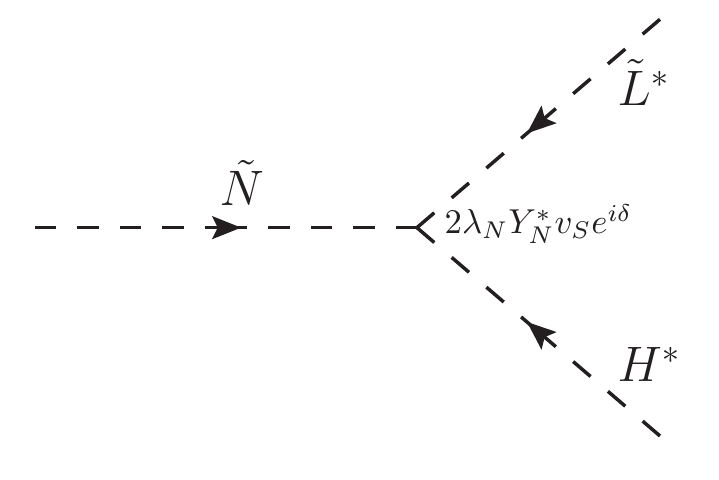}\includegraphics[height=3cm,width=4cm]{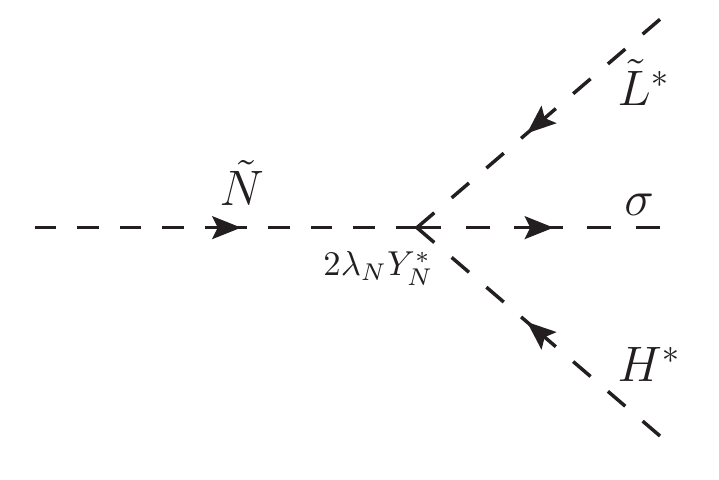}
\caption{{2-body and 3-body decay diagrams of a singlet sneutrino, $\tilde{N}$.}}
\label{fig:FD-SN}
\end{center}
\end{figure}

Irrespective of the mass of $\sigma$ relative to the mass of $\tilde{N}$, the sneutrino can decay into leptonic (sleptonic) and Higgs (Higgsino) final particles. The CP asymmetry generated from such a scenario was calculated in~\cite{Grossman:2003jv, DAmbrosio:2003nfv}. However, if $M_1\gg m_\sigma$, the {3-}body decay channel shown in the last Feynman diagram of Fig. \ref{fig:FD-SN} opens up and it leads to interesting consequences for soft leptogenesis as we show below.

For the snuetrino system, upto leading order in the off-diagonal terms\footnote{Considering $\kappa$ and $A_\lambda$ much smaller compared to $\lambda_N$ and $M_1$ respectively as will be justified later.}, the mass matrix can be calculated from the {squared mass} matrix in Eq.~(\ref{eq:M2N}):
\begin{align}
{\cal M}_{\tilde{N}} = M_1\begin{bmatrix}
1 & \frac{\lambda_N\kappa v^2_{S} e^{2i\delta}}{2 M_1^2}+\frac{A_\lambda \lambda_N v_{S} e^{-i\delta}}{2 M_1^2} \\
\\
\frac{\lambda_N\kappa v^2_{S} e^{-2i\delta}}{2 M_1^2}+\frac{A_\lambda \lambda_N v_{S} e^{i\delta}}{2 M_1^2} & 1
\end{bmatrix}.\label{eq:mass}
\end{align}
\begin{figure}[h]
\begin{center}
\includegraphics[height=3.5cm,width=4.2cm]{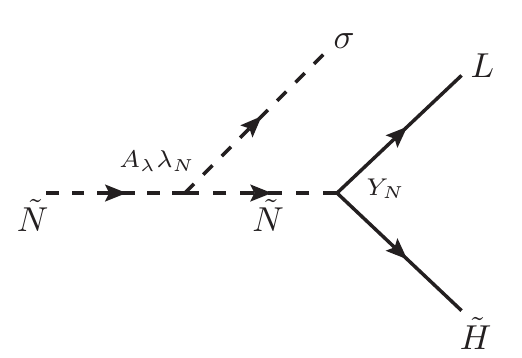}
\includegraphics[height=3.5cm,width=4.2cm]{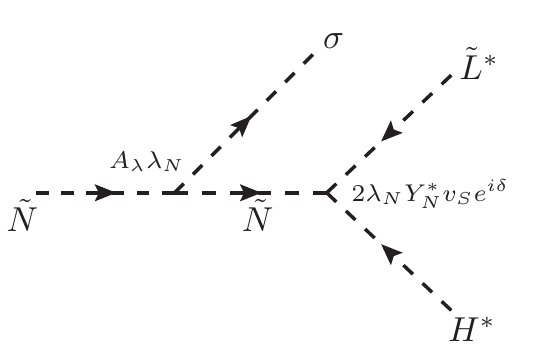}\includegraphics[height=3.5cm,width=4.2cm]{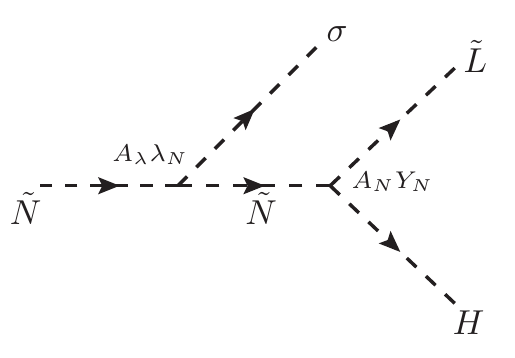}
\caption{{3-body decay diagrams of singlet sneutrino, $\tilde{N}$ at higher order.}}
\label{fig:FD-SN1}
\end{center}
\end{figure}
The decay rate matrix can be written from the Eq.~(\ref{eq:interaction}). It contains both diagonal as well as off-diagonal terms because $\tilde{N}$ can decay into particle as well as anti-particle final states.  
\begin{eqnarray}
\Gamma_{\tilde{N}} &=& \Gamma_1\begin{bmatrix}
1+\alpha+\beta+\frac{4\lambda_N^2 v^2_{S}}{M_1^2} + \frac{|A_N|^2}{M_1^2} & \frac{4\lambda_N v_{S} e^{-i\delta} A_N^*}{M_1^2} \\ \\
\frac{4\lambda_N v_{S} e^{i\delta} A_N}{M_1^2} & 1+\alpha+\beta+\frac{4\lambda_N^2 v^2_{S}}{M_1^2} + \frac{|A_N|^2}{M_1^2}
\end{bmatrix},\label{eq:gamma}\\ \nonumber 
\end{eqnarray}
with,
\begin{align}
&\Gamma_1 =\frac{|Y_{N}|^2 M_1}{8\pi}{.}
\end{align}
{The parameter $\alpha$ and $\beta$ in $\Gamma_{\tilde{N}}$ matrix are associated with the 3-body decay contributions of $\tilde{N}$ given by,
\begin{align}
 \alpha~ \simeq&~ \frac{\lambda_N^2}{\pi^2}\Bigg[\frac{M_1(M_1^2+4 m_\sigma^2)^{1/2}}{2M_1^2}-2\frac{m_\sigma^2}{M_1^2}\log\left\{\frac{(M_1^2+4m_\sigma^2)^{1/2}+M_1}{2m_\sigma} \right\}+\frac{|A_\lambda|^2v_S^2}{2 M_{1}^4}{\rm ~log}\left(\frac{m_N}{2 m_\sigma}\right)\nonumber\\
 &+\frac{A_\lambda v_S \cos \delta}{2 M_1^2}\left\{-1+\frac{m_\sigma}{2M_1}+\frac{m_\sigma^2}{4 M_1^2}{\rm~log}\left(\frac{2M_1m_\sigma-m_\sigma^2}{M_1^2-m_\sigma^2}\right)\right\}\Bigg],\label{eq:alpha}\\
 \beta~\simeq&~ \frac{\lambda_N^2|A_\lambda|^2}{2\pi^2M_1^2}\Bigg[{\rm log}\left\{\frac{1}{8}+\frac{|A_N|^2}{M_1^2}\right\}-\frac{1}{8}\Bigg]\label{eq:beta}.
\end{align}
The logarithm in $\beta$ can be expanded for the soft term $A_N$ at least an order of magnitude smaller than $M_1$ (this can be assumed without any loss of generality as later to draw our conclusions and compare with MSSM soft-leptogenesis in \cite{DAmbrosio:2003nfv, Fong:2011yx, Grossman:2003jv}, we will be restricting to leading order in soft-terms) to give:
\begin{equation}
\beta \simeq  \frac{\lambda_N^2|A_\lambda|^2}{2\pi^2M_1^2} \Bigg[ \frac{A_N^2}{M_1^2}-\frac{3}{4}\Bigg]
\end{equation}
Under the same approximation used for $\beta$ and with $M_1\gg m_\sigma$, $\alpha$ becomes:
\begin{align}
\alpha \simeq \frac{\lambda_N^2}{\pi^2}\Bigg[\frac{1}{2}-\frac{A_\lambda v_S \cos\delta}{2M_1^2} \Bigg]
\end{align}
It is important to note that as long as the soft term $A_\lambda$ is sufficiently small compared to $M_1$ (at least by an order of magnitude) or the phase $\pi/2 < \delta < 3\pi/2$, $\alpha\neq 0$.

There are 4 terms in $\alpha$ of Eq.(\ref{eq:alpha}) which represents the contribution of $\tilde{N}\rightarrow \sigma \tilde{L}^* H^*$ decay,  where the first two terms come from the 3-body decay arising from the point interaction in Fig. \ref{fig:FD-SN} while the third term comes from the second diagram of Fig. \ref{fig:FD-SN1}. The last term in $\alpha$ comes from the interference of these two diagrams.  The $\beta$ in Eq.(\ref{eq:beta}) appears in the decay matrix $\Gamma_{\tilde{N}}$ due to the presence of first and the third diagrams of Fig. \ref{fig:FD-SN1}. It is pertinent to note here that $\beta$ is at least quadratic in the soft terms $A_\lambda$ and $A_N$. One could in principle have higher order diagrams relating to 4 or more decay products. However, all such diagrams will be suppressed heavily by higher powers of $A_\lambda/M_1$.

 The solutions for the time evolution of $\tilde{N}$ and $\tilde{N}^*$ come from the Schrodinger like equation
\begin{equation}
{\cal H}\psi = i\frac{d\psi}{dt},
\end{equation}
where $\psi=\{\tilde{N},\tilde{N}^*\}^T$. The solutions are obtained as,
\begin{eqnarray}
\tilde{N}(t) &=& e^{-iat}\left[\tilde{N}_0 \cos\left(\frac{p}{q} \;bt\right)-i\tilde{N}^*_0 \frac{q}{p}\sin\left(\frac{p}{q}\; bt\right)\right],\\
\tilde{N}^*(t) &=& e^{-iat}\left[\tilde{N}^*_0 \cos\left(\frac{p}{q} \;bt\right)-i\tilde{N}_0 \frac{p}{q}\sin\left(\frac{p}{q}\; bt\right)\right],
\end{eqnarray}
where $\tilde{N}_0,\tilde{N}^*_0$ are the field values at $t=0$ and
\begin{eqnarray}
a&=&{({\cal M}_{\tilde{N}})}_{11}-i\frac{{(\Gamma_{\tilde{N}})}_{11}}{2}={({\cal M}_{\tilde{N}})}_{22}-i\frac{{(\Gamma_{\tilde{N}})}_{22}}{2},\\[0.1cm]
b&=&{({\cal M}_{\tilde{N}})}_{12}-i\frac{{(\Gamma_{\tilde{N}})}_{12}}{2},\\[0.1cm]
\left(\frac{p}{q}\right)^2&=&\frac{{({\cal M}_{\tilde{N}})}_{12}^*-i\frac{{(\Gamma_{\tilde{N}})}_{12}^*}{2}}{{({\cal M}_{\tilde{N}})}_{12}-i\frac{{(\Gamma_{\tilde{N}})}_{12}}{2}}.\\ \nonumber
\end{eqnarray}
Let's define $\Delta M = M_+ - M_-$ and $\Delta \Gamma_{\tilde{N}} = \Gamma_+ -\Gamma_-$ and $Q=\frac{p}{q}b$. Then if $\Gamma_1^2\ll M_1^2$ which happens when $Y_{N}\ll 1$ (typically of $\mathcal{O}(10^{-4})$ to satisfy neutrino mass bounds), we can write
\begin{eqnarray}
2 \;{\rm Re} (Q)\simeq \Delta M &=& \frac{\lambda_N v_{S}}{M_1}\left[A_\lambda^2 + \kappa^2 v^2_{S} +2A_\lambda v_{S}\kappa \cos(3\delta)\right]^{1/2}, \label{eq:DeltaM}\\
-4\; {\rm Im} (Q)\simeq \Delta \Gamma_{\tilde{N}} & = & \frac{2Y_N^2\lambda_N^2 v^2_{S}}{\pi^2M_1^2\Delta M}\left\{\kappa v_{S}(\cos (3\delta){\rm Im}A_N+\sin (3\delta){\rm Re}A_N)+A_\lambda{\rm Re} A_N\right\}.\label{eq:DeltaGamma}
\end{eqnarray}
If $\Delta\Gamma_{\tilde{N}}\ll \Delta M$ as well, the argument of the trigonometric functions becomes {$\Delta M t/2$} such that we can write:
\begin{eqnarray}
\tilde{N}(t) &=& g_1 \tilde{N}_0 +\frac{q}{p} g_2 \tilde{N}^*_0, \label{eq:Nmix}\\
\tilde{N}^*(t) &=& g_1 \tilde{N}^*_0 +\frac{p}{q} g_2 \tilde{N}_0, \label{eq:NsMix}
\end{eqnarray}
where{}
\begin{eqnarray}
g_1 &=& e^{-iM_1 t} \; \exp\left[-\frac{\Gamma_1}{2}\left(1+\alpha+\beta+\frac{4\lambda_N^2 v^2_{S}}{M_1^2}+\frac{|A_N|^2}{M_1^2}\right)t\right] \cos\left[\frac{\Delta M t}{2}\right], \\
\nonumber\\
g_2 &=& -i e^{-iM_1 t} \; \exp\left[-\frac{\Gamma_1}{2}\left(1+\alpha+\beta+\frac{4\lambda_N^2 v^2_{S}}{M_1^2}+\frac{|A_N|^2}{M_1^2}\right)t\right]\sin\left[\frac{\Delta M t}{2}\right].
\end{eqnarray}
The Eqs. (\ref{eq:Nmix}) and (\ref{eq:NsMix}) are substituted back in Eq. (\ref{eq:interaction}) for calculating the CP asymmetry factor $\varepsilon$ which is defined as the ratio of the difference between the decay rates of $\tilde{N}$ and $\tilde{N}^*$ into final state particles with lepton number $+1$ and $-1$ to the sum of all the decay rates{,} {\it i.e.}{,}
\begin{align}
\varepsilon = \frac{\sum_f \int_0^\infty dt\;\left[\Gamma(\tilde{N}(t)\rightarrow f)+\Gamma(\tilde{N}^*(t)\rightarrow f)-\Gamma(\tilde{N}(t)\rightarrow \bar{f})-\Gamma(\tilde{N}(t)\rightarrow \bar f)\right]}{\sum_f \int_0^\infty dt \; \left[\Gamma(\tilde{N}(t)\rightarrow f)+\Gamma(\tilde{N}^*(t)\rightarrow f)+\Gamma(\tilde{N}(t)\rightarrow \bar f)+\Gamma(\tilde{N}^*(t)\rightarrow \bar f)\right]},
\end{align}
{where $f,~\bar f$ are the final states with lepton number $+1$ and $-1$, respectively.}
This then gives us the following CP asymmetry parameter:
\begin{align}
\varepsilon=\frac{\int_0^\infty\;dt\;\frac{|g_2|^2|Y_{N}|^2 M_1}{8\pi}\left(\left|\frac{q}{p}\right|^2-\left|\frac{p}{q}\right|^2\right) \left[1+\frac{|A_N|^2}{M_1^2}-\frac{4\lambda_N^2 v^2_{S}}{M_1^2}-\alpha+\beta\right]}{\int_0^\infty\;dt\;\frac{|Y_{N}|^2M_1}{4\pi}\left(1+\frac{|A_N|^2}{M_1^2}+\frac{4\lambda_N^2 v^2_{S}}{M_1^2}+\alpha+\beta\right)\;\left[|g_1|^2+\frac{|g_2|^2}{2}\left(\left|\frac{q}{p}\right|^2+\left|\frac{p}{q}\right|^2\right)\right]}.
\end{align}
The decay rates at the tree{-}level itself for final states with lepton numbers $\pm 1$ are different because the factor $\left|\frac{p}{q}\right|\neq 1$ as it is not a hermitian quantity. This requires non-zero off-diagonal terms to be present in the mass matrix as well as the decay rate matrix of the system with atleast one of them being complex.

The sum and difference of the ratios $|q/p|^2$ and $|p/q|^2$ can be written as:
\begin{eqnarray}
\left|\frac{q}{p}\right|^2-\left|\frac{p}{q}\right|^2 & = & -2\left(\frac{y^2}{x^2-y^2}\right)^{1/2},\\
\left|\frac{q}{p}\right|^2+\left|\frac{p}{q}\right|^2 & = & 2\left(\frac{x^2}{x^2-y^2}\right)^{1/2},
\end{eqnarray}
where
\begin{eqnarray}
x &=& \frac{\lambda_N^2\kappa^2 v^4_{S}}{4}+\frac{A_\lambda^2 \lambda_N^2 v^2_{S}}{4}+\frac{4\Gamma_1^2 \lambda_N^2 v^2_{S} |A_N|^2}{M_1^2}+\frac{\lambda_N^2 A_\lambda \kappa v^3_{S} \cos(3\delta)}{2}, \\
y &=& \frac{2\Gamma_1 A_\lambda \lambda_N^2 v^2_{S}{\rm Im}A_N}{M_1}+\frac{4\lambda_N^2\kappa v^3_{S} \Gamma_1}{M_1}{\Big(}\cos(3\delta) {\rm Im}A_N + \sin(3\delta){\rm Re}A_N{\Big)}.
\end{eqnarray}
Keeping terms upto the leading order in $\Gamma_{1}$, we find that the sum of the ratios is $\simeq 2$ while the difference is twice the values of $y/(x^2-y^2)^{1/2}$ with,
\begin{equation}
\left(\frac{y^2}{x^2-y^2}\right)^{1/2} =\frac{|Y_{N}|^2{\Big[}A_\lambda {\rm Im}A_N+ \kappa v_{S}\{\cos(3\delta) {\rm Im}A_N + \sin(3\delta){\rm Re}A_N\}{\Big]}}{\pi {\Big[}\kappa^2 v^2_{S}+A_\lambda^2+2 A_\lambda \kappa v_{S} \cos(3\delta){\Big]}}\label{eq:redef1}
\end{equation}
such that the final CP asymmetry can be written as:
\begin{align}
 \varepsilon=-\frac{\Delta M^2}{2(\Gamma^2+\Delta M^2)}\frac{\left[1+\frac{|A_N|^2}{M_1^2}-\frac{4\lambda_N^2 v^2_{S}}{M_1^2}-\alpha+\beta\right]}{\left[1+\frac{|A_N|^2}{M_1^2}+\frac{4\lambda_N^2 v^2_{S}}{M_1^2}+\alpha+\beta\right]}\left(\frac{y^2}{x^2-y^2}\right)^{1/2}{,}\label{eq:asymmerty1}
\end{align}
where $\Gamma$ is defined as $\Gamma=\Gamma_1 \left[1+\frac{|A_N|^2}{M_1^2}+\frac{4\lambda_N^2 v^2_{S}}{M_1^2}+\alpha+\beta\right]$.
From Eqs.~(\ref{eq:redef1}) and (\ref{eq:asymmerty1}) it is evident that there is {a} non-zero CP asymmetry even with real $A_N$, provided $\delta$ is sufficiently large.

We also note that while the usual soft leptogenesis done in {the} MSSM~\cite{Grossman:2003jv, DAmbrosio:2003nfv}, to the leading order in soft terms ($M^2\ll M_1^2, A_N^2\ll M_1^2$),  necessarily requires thermal phase space factors for the final state bosons ($c_B$) and fermions ($c_F$) to have $\varepsilon\propto \Delta_{BF}=\frac{c_B-c_F}{c_B+c_F}$, we get an asymmetry even {without thermal mass corrections to the decay products at the leading order in soft terms.} This happens because of the presence of $\sigma$ in the model which facilitates a 3-body decay of $\tilde{N}$ which is not cancelled by the other terms. If we did not have this, at leading order in soft terms, $1-\frac{4\lambda^2_N v^2_{S}}{M_1^2}=0$ and there would be no asymmetry without thermal mass corrections. At the leading order in the soft terms ($A_\lambda\ll M_1^2\Rightarrow \beta$ is negligible), $\varepsilon$ is 
\begin{align}
 \varepsilon=\frac{\Delta M^2}{2(\Gamma^2+\Delta M^2)}\frac{\alpha}{\left(2+\alpha\right)}\left(\frac{y^2}{x^2-y^2}\right)^{1/2}{.}\label{eq:asymLead}
\end{align}
It is clear from Eq. (\ref{eq:asymLead}) that the only contribution coming from diagrams of Fig. \ref{fig:FD-SN1} are coming from the cross term of the second diagram with the last diagram of Fig. \ref{fig:FD-SN} through $\alpha$. Even in $\alpha$, the dominant contribution comes through the 3-body decay by the vertex interaction of Fig. \ref{fig:FD-SN}. Therefore it is possible to successfully generate non{-}zero lepton asymmetry from $\tilde{N}-\tilde{N}^*$ system at the tree{-}level without using thermal phase space factors for bosonic and fermionic final states. We discuss this more and give some numerical estimates in Sec.~\ref{sec:results} for relevant parameters. For the moment, let's consider the decays of $\sigma$.
\vspace{-0.6cm}
\subsection{$\sigma$ decays}\label{sec:sigmadecay}
\vspace{-0.3cm}
Unlike the $\tilde{N}$ decays where most of the decay products were massless, the final products of $\sigma$ decay are massive, {as shown in Fig.~\ref{fig:FD-sigma}}. This creates two possible decay modes of $\sigma$ according to the condition satisfied.
\begin{figure}[t!]
\begin{center}
\includegraphics[height=3.5cm,width=4.5cm]{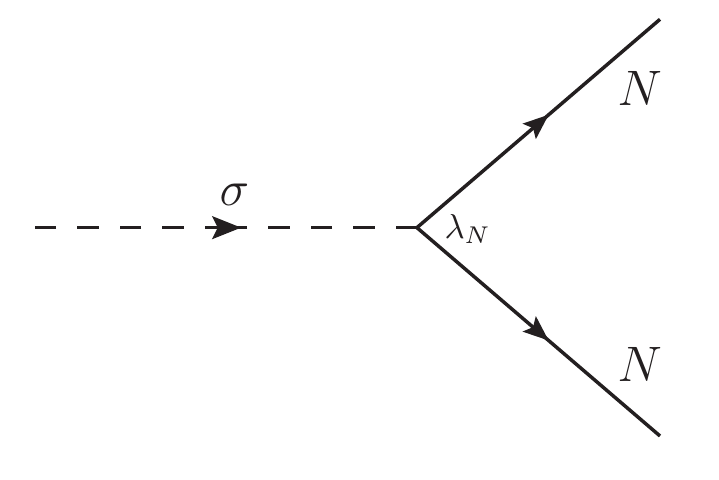}
\includegraphics[height=3.5cm,width=4.5cm]{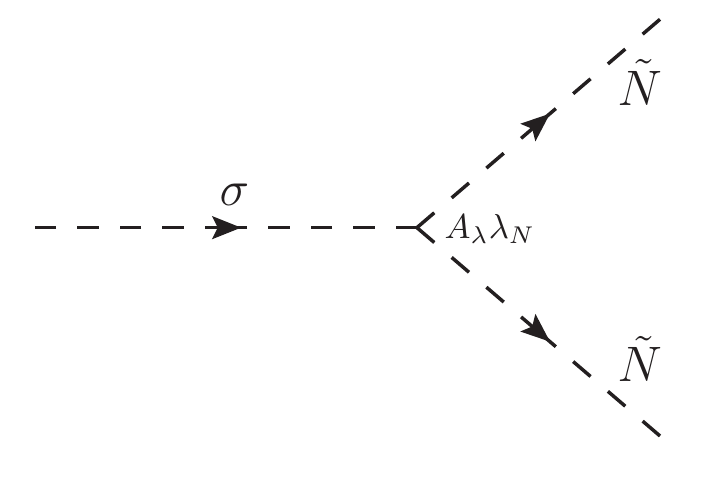}
\caption{{Decay diagrams of a singlet scalar $\sigma$.}}
\label{fig:FD-sigma}
\end{center}
\end{figure}
\begin{enumerate}
\item $\sigma$ decays to $NN$ and $\tilde{N}\tilde{N}$. This happens when $m_\sigma^2 > 4 M_1^2 ${,}
\item $\sigma$ decays only to $NN$. This happens when $16\lambda_N^2 v^2_{S} = 4m_N^2 < m_\sigma^2 < 4M_1^2$.\end{enumerate}
The mass matrix and the $\Gamma$ matrix of the  $\sigma-\sigma^*$ system are respectively:
\begin{eqnarray}
M_\sigma &=& m_\sigma \begin{bmatrix}
1 & \frac{\kappa^2 v^2_{S}\,e^{2i\delta}+A_\kappa^* \kappa  v_{S} e^{-i\delta}}{m_\sigma^2} \\ \\
\frac{\kappa^2 v^2_{S}\,e^{-2i\delta}+A_\kappa \kappa  v_{S} e^{i\delta}}{m_\sigma^2} & 1
\end{bmatrix},~~~~~\Gamma_\sigma = \Gamma_{\sigma,1}\begin{bmatrix}
\Theta  & 0 \\
0 & \Theta 
\end{bmatrix},
\end{eqnarray}
where 
\begin{align}
\Gamma_{\sigma,1} &= \frac{\lambda_N^2}{32\pi m_\sigma^2}, \\[0.2cm]
\Theta &= \left\lbrace \begin{array}{cc}
\left(m_\sigma^2-4m_N^2 \right)^{3/2}+A_\lambda^2 \left(m_\sigma^2 - 4 M_1^2 \right)^{1/2} & \textrm{for case 1},\\ \\
\left(m_\sigma^2-4m_N^2 \right)^{3/2} & \textrm{for case 2}.
\end{array}\right.
\end{align}
A non-relativistic Hamiltonian can be defined following Eq.~(\ref{eq:Hamiltonian}). Immediately it can be seen that because of the absence of an off-diagonal term in the decay rate matrix of $\sigma$, the ratio corresponding to $(p/q)^2$ of $\tilde{N}$ decay,
\begin{align}
\left(\frac{s}{r}\right)^2=\frac{(M_\sigma)_{12}^*-i\frac{(\Gamma_{\sigma}^*)_{12}}{2}}{(M_\sigma)_{12}-i\frac{(\Gamma_{\sigma})_{12}}{2}}=\frac{(M_\sigma)_{12}^*}{(M_\sigma)_{12}}{.}
\end{align}
If we solve for the evolution of the $\sigma-\sigma^*$ system, the CP asymmetry parameter computed exactly analogously to the $\tilde{N}-\tilde{N}^*$ system will be zero because of exact cancellation between the ratios $|r/s|^2$ and $|s/r|^2$. Thus 
\begin{equation}
\varepsilon_\sigma = 0.
\end{equation}

\section{General Constraints}
\label{sec:consts}
The CP asymmetry in this model depends on a lot of parameters. However, we can constrain some of them by various considerations. In deriving the following general constraints, we take $M_1\gg M \Rightarrow M_1 \simeq 2\lambda_N v_{S}$, $M_1\gg m_\sigma$ and $A_\lambda\simeq \kappa v_{S}$ for reasons that will become clear later in Sec.~\ref{sec:results}.
\begin{itemize}
 \item The condition of out-of-equilibrium decay at $T=M_1$ is given by comparing the decay rate of the $\tilde{N}$ with the Hubble parameter at $T=M_1$:
 \begin{align}
  \Gamma\lesssim H(T=M_1) = \sqrt{\frac{8\pi^3 g_s}{90}} \frac{M_1^2}{m_{Pl}}, \label{eq:gammaEQ}
 \end{align}
 where $\Gamma$ is the diagonal component of Eq.~(\ref{eq:gamma}). Substituting it in Eq.~(\ref{eq:gammaEQ}) and neglecting the contribution of $\beta$ we get
 \begin{align}
  \frac{|A_N|^2}{M^2_1}\lesssim \frac{13 \pi \sqrt{g_*} M_1}{Y^2_{N} m_{Pl}}-2-\alpha.\label{eq:limitOut1}
 \end{align}
For $M_1 \gg m_\sigma$, $\alpha\approx \mathcal{O}(10^{-2})$ and we may write Eq.~(\ref{eq:limitOut1}) as
 \begin{align}
  \frac{|A_N|^2}{M^2_1}\lesssim \frac{13 \pi \sqrt{g_*} M_1}{Y_N^2 m_{Pl}}-2.\label{eq:limitOut2}
 \end{align}

 \item The way we derived the CP asymmetry requires well separated states~\cite{DAmbrosio:2003nfv}, i.e., $\Gamma \ll \Delta M$ as well as $\Delta\Gamma\ll \Delta M$ as stated before. This gives us two self-consistent limits: 
 \begin{align}
  &~~~~~~~~~~~~~\frac{|A_N|^2}{M_1^2} \ll   ~\frac{8 \pi \kappa v_{S}}{Y_N^2 M_1}-2, \\
  &\cos(3\delta) {\rm Im}A_N + \left(\sin(3\delta)+1\right) {\rm Re}A_N \ll   ~\frac{2\pi^2}{Y_N^2} \kappa v_{S}. 
 \end{align}
 \item Neutrino mass upper limits ($m_\nu\lesssim 0.1$ eV~\cite{Aghanim:2018eyx, Tanabashi:2018oca}) put constraints on the Yukawa coupling strength $Y_{N}$ and the mass of the {RH} (s)neutrino. 
 \begin{align}
  \frac{Y^2_{N}}{\lambda_N v_{S}}\lesssim 6.6\times 10^{-15} ~{\rm GeV},\label{eq:Ylimit}
 \end{align}
 
\item Electric dipole moment {(EDM)} calculations can constrain the CP violating phases that appear in the vevs of the two Higgs doublets and the scalar singlet $S$. In {Ref.}~\cite{Huitu:2012rd}, they show that in principle $\delta_u$ (the phase in the vev of $H_u$ should we go below the EW scale) and $\delta$ could be large as long as the relative phase is kept small. For more details about the EDM constraints on the NMSSM, see~\cite{King:2015oxa}.
 
\end{itemize}

\subsection{A simpler form for $\varepsilon$}
We can write a simpler form for the CP asymmetry by using the approximations made and the general relationships between various parameters given above. The set of parameters governing $\varepsilon$ are:
\begin{align}
 \Big\{m_S,~M,~\kappa,~\lambda_N,~v_S,~Y_{N},~A_\lambda,~{\rm ReA_N},~{\rm Im} A_N,~\delta\Big\}.
\end{align}
We choose the soft masses $M,~m_{S}\sim \mathcal{O}(1)$~TeV and {$v_S$ to be of ${\cal O}(10^7)$~GeV} with $\lambda_N\sim\mathcal{O}(1)$. This along with $\kappa\ll \lambda_N$ means both $M,m_\sigma\ll 2\lambda_N v_{S}\simeq M_1$. 
{Therefore, from Eq.~(\ref{eq:Ylimit}), one can put an upper limit on $Y_N${,} {\it i.e.}{,} $Y_{N}\lesssim\mathcal{O}(10^{-4})$}. With these choices and approximations the form of $\varepsilon$ can be simplified to
\begin{align}
\varepsilon \simeq \frac{1}{2\pi}\left(\frac{\alpha}{2+\alpha}\right)\times \frac{Y_N^2{\Big[}A_\lambda {\rm Im}A_N + \kappa v_S\{\cos(3\delta) {\rm Im}A_N+\sin(3\delta){\rm Re}A_N\}{\Big]}}{A_\lambda^2+\kappa^2 v_S^2+2A_\lambda \kappa v_S\cos(3\delta)}{.} \label{eq:asymSimp}
\end{align}
\section{Results and discussions}\label{sec:results}
To obtain the baryon asymmetry of the Universe, $\eta_B${,} we solve the simultaneous Boltzmann equations for the $\tilde{N}$ number density, $N_{\tilde{N}}$, and the $B-L$ number density, $N_{B-L}${,} which are as follows~\cite{Buchmuller:2004nz,Garayoa:2009my,HahnWoernle:2009qn}:
\begin{eqnarray}
\frac{dN_{\tilde{N}}}{dz}&=& -K_{\tilde{N}} z (N_{\tilde{N}}-N_{\rm eq}) \frac{\kappa_1(z)}{\kappa_{2}(z)},\\
\frac{dN_{B-L}}{dz}&=& -\varepsilon K_{\tilde{N}} z(N_{\tilde{N}}-N_{\rm eq})\frac{\kappa_1(z)}{\kappa_2(z)}-\frac{1}{4}K_{\tilde{N}} z^3 \kappa_1(z) N_{B-L},
\end{eqnarray}
where $K_{\tilde{N}}=\frac{\Gamma}{H(z=1)}$ is the Hubble parameter at $z=1$ with $z=\frac{M_1}{T}$ {and} $N_{\rm eq}$ is the equilibrium number density of $\tilde{N}$. They take the following forms:
\begin{eqnarray}
H(z=1)&=& \sqrt{\frac{8\pi^3 g_s}{90}} \frac{M_1^2}{m_{Pl}},\\
N_{\rm eq} &=& \kappa_2(z)\frac{z^2}{2},
\end{eqnarray}
with $m_{Pl}=1.22\times 10^{19}$~{GeV} being the Planck mass and $g_s$ is the number of relativistic degrees of freedom in {the} NMSSM which we take $\approx 225$ except for the $\tilde{N}$ which is non-relativistic. In writing the Boltzmann equation for $B-L$ number density, we neglect the $\Delta L=2$ scattering processes for washout and assume it is dominated mostly by inverse decays. The contribution to washout from the scattering processes is small because we are in the weak washout regime with $K_{\tilde{N}}\lesssim 1$. The final $B-L$ number density thus created, $N_{B-L}^f$ then converts to the baryon asymmetry by the sphaleron processes such that the ratio of the baryon number density to the photon number density, $\eta_B${,} is:
\begin{equation}
\eta_B = \frac{3}{4}\frac{g_*^0}{g_*}a_{{\rm sph}} N_{B-L}^f,
\end{equation}
where $g_*\simeq g_s\simeq 225$, $g_*^0$ is the effective number of relativistic degrees of freedom at recombination and $a_{{\rm sph}}$ is the sphaleron conversion factor. Since we will solve the Boltzmann equations numerically, we use the complete form of $\varepsilon$ given in Eq.~(\ref{eq:asymmerty1}){.}

\begin{figure}[h]
\begin{center}
\includegraphics[height=7cm,width=12cm]{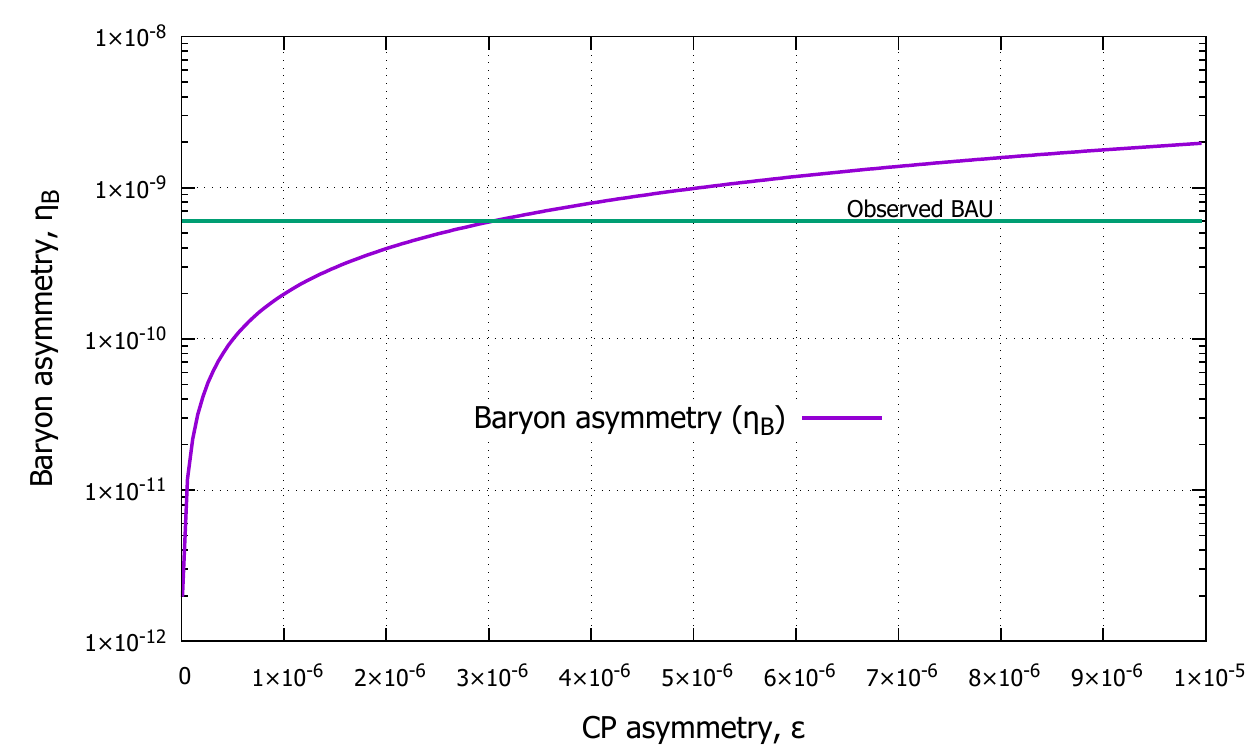}
\caption{The dependence of the baryon asymmetry on the CP asymmetry. In this figure, we keep $Y_{N}=10^{-4.5}, ~\lambda_N = 0.9, ~M = 10^4~{\rm GeV}, ~v_{S} = 10^{6.5}~{\rm GeV}, ~|A_N|=10^{6.5}~{\rm GeV}, ~m_\sigma = 10^2~{\rm GeV}$. The variation in $\varepsilon$ is brought on by not fixing $A_\lambda, \kappa, \delta$. We also take equilibrium initial condition for $\tilde{N}$ abundance. Starting with zero initial equilibrium for $\tilde{N}$ does not change the result.}
\label{fig:BAU}
\end{center}
\end{figure}
\hspace{0.2cm}
In Fig.~\ref{fig:BAU}, we show the typical value of the CP asymmetry that satisfies the observed baryon asymmetry of the Universe. It turns out that we need $\varepsilon \simeq \mathcal{O}(10^{-6})$ to get the correct observed baryon asymmetry while satisfying neutrino mass bounds. This value of $\varepsilon$ is similar to the one obtained by other vanilla leptogenesis scenarios in the weak washout regime. The only difference is that usual leptogenesis occurs with decays at the loop{-}level interfering with tree{-}level decays due to complex Yukawa couplings that violate CP. In the soft leptogenesis, the Yukawa parameter could very well remain real as the source of CP asymmetry lies elsewhere -- in the time-varying mixing between $\tilde{N}$ and $\tilde{N}^*$ states due to the complex nature of $\langle S\rangle$ and the soft SUSY breaking parameters.

\subsection{Case 1: $\delta$ is near zero}\label{sec:nonres}
To get $\varepsilon\simeq \mathcal{O}(10^{-6})$ we fix the values of the following parameters in line with the earlier approximations and constraints:
\begin{align}
 \delta=0.3, M=1~{\rm TeV},~ m_S=1~ {\rm TeV},~\lambda_N=1.
\end{align}
{For simplicity}, we also assume $A_\lambda\simeq\kappa v_S$ (such that $\alpha\approx \frac{\lambda_N^2}{2\pi}$) in finding the correct set of values for other parameters. Using these, we show the relation between $A_N$ and $\kappa$ for different values of $Y_{N}$ and $v_S$ (satisfying the SM neutrino mass bounds) in Figs.~\ref{fig:ReAnAl} and \ref{fig:ImAnAl} as contour plots in ${\log}~\varepsilon$. 
\begin{figure}[h]
\begin{center}
\includegraphics[height=7cm,width=8.1cm]{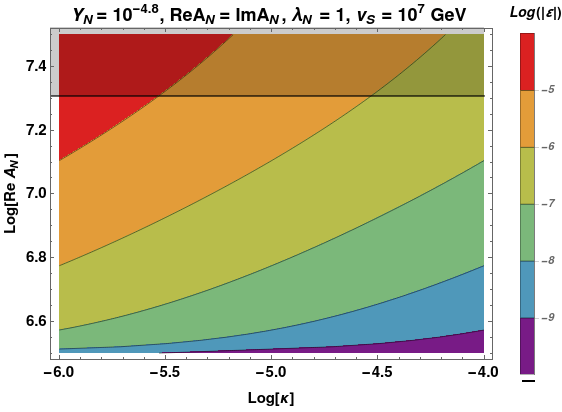}
\includegraphics[height=7cm,width=8.1cm]{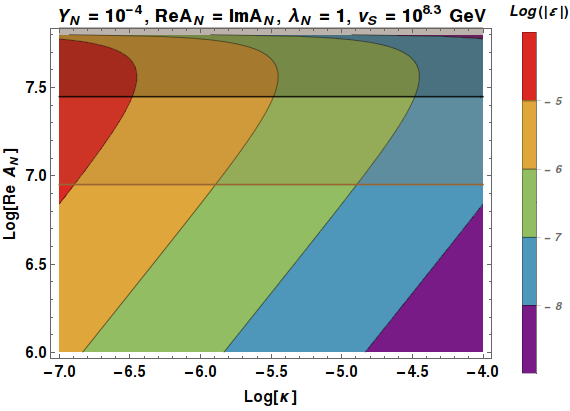}
\caption{The contour plots of $\varepsilon$ in ${\rm Re} A_N-\kappa$ plane considering $A_N$ complex. The left panel shows only the out-of-equilibrium bound ($\Gamma\simeq H(z=1)$, in horizontal black solid line) with a generic $A_N$. In the right panel the horizontal brown solid line marks the leading order approximation in the soft-term $A_N$ ($A_N^2/M_1^2\simeq 10^{-3}$), while the black line is the out-of-equilibrium bound. All logarithms are to the base 10.}
\label{fig:ReAnAl}
\end{center}
\end{figure}

In Fig.~\ref{fig:ReAnAl} we take a complex $A_N$ with ${\rm Im}A_N = {\rm Re}A_N$ and vary $\kappa$ and Re$A_N$ for $Y_N=10^{-4.8}, v_S = 10^7$~GeV (left panel) and $Y_{N}=10^{-4}, v_S=10^{8.3}$~GeV (right panel). As can be seen from the left figure, we need large $A_N$ ($\gtrsim M_1$) to satisfy correct order of $\varepsilon$ while ensuring that $\Gamma\lesssim H(z=1)$. However from the right panel, it is clear that sufficient CP asymmetry can be created even at leading order in $A_N$ without using thermal phase space factors. 

We keep $A_N$ real in Fig.~\ref{fig:ImAnAl} and vary $\kappa$ and ${\rm Re}A_N$ for similar values of $Y_{N}$ and $v_S$ as before. It's clear that non{-}zero asymmetry can be created even with real $A_N$ as long as $\delta\neq n\pi$ {($n \in \mathbb{Z}$)}. However compared to Fig.~\ref{fig:ReAnAl}, we find that the bounds on $A_N$ and $\kappa$ in Fig.~\ref{fig:ImAnAl} are stronger as there is much less parameter space available satisfying $\varepsilon\gtrsim \mathcal{O}(10^{-6})$. Both Figs.~\ref{fig:ReAnAl} and \ref{fig:ImAnAl} satisfy neutrino mass bounds and the conditions of $\Gamma,\Delta\Gamma \ll \Delta M$.
\begin{figure}[h]
\begin{center}
\includegraphics[height=7cm,width=8.1cm]{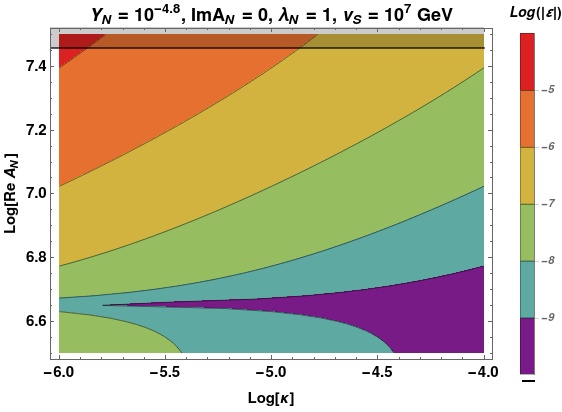}
\includegraphics[height=7cm,width=8.1cm]{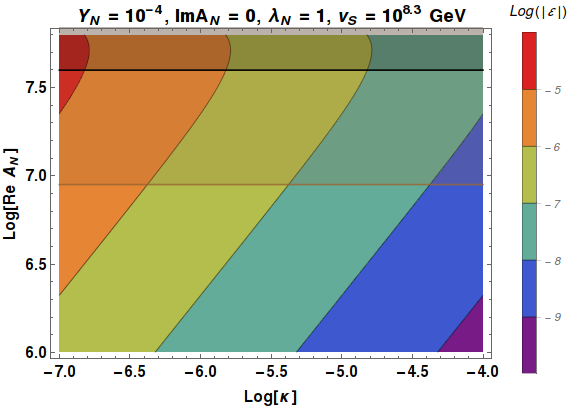}
\caption{The contour plots of $\varepsilon$ in ${\rm Re} A_N-\kappa$ plane considering a real $A_N$. The left panel shows only the out-of-equilibrium bound ($\Gamma\simeq H(z=1)$, in horizontal black solid line) with a generic $A_N$. In the right panel the horizontal brown solid line marks the leading order approximation in the soft-term $A_N$ ($A_N^2/M_1^2\simeq 10^{-3}$), while the black line is the out-of-equilibrium bound. All logarithms are to the base 10.}
\label{fig:ImAnAl}
\end{center}
\end{figure}

\subsection{Case 2: $\delta=\pi$}
If the phase of the vev of $S$ is large, specially at $\delta =\pi$, we get a resonance behaviour in $\varepsilon$ at $A_\lambda \simeq \kappa v_S$. In the limit of $\delta\rightarrow \pi$, the CP asymmetry parameter of Eq.~(\ref{eq:asymSimp}) can be written as:
\begin{align}
\varepsilon\big|_{\delta=\pi} = \frac{1}{2\pi}\left(\frac{\alpha+\beta}{2+\alpha+\beta}\right)\frac{Y_N^2 \;{\rm Im}A_N}{|A_\lambda-\kappa v_S|}{.}\label{eq:asymDpi}
\end{align}
Eq.~(\ref{eq:asymDpi}) also justifies assuming $A_\lambda \simeq \kappa v_S$ to derive the general constraints on the various parameters in Sec.~\ref{sec:consts}. The behaviour of $\varepsilon$ versus $A_\lambda$ is shown in Fig.~\ref{fig:res}. For the plot, we take $M=m_S=1$~TeV. The values of the other relevant parameters are shown in the figure itself. Since $\delta=\pi$, there is no contribution from the real part of $A_N$ in $\varepsilon$. This means that $A_N$ necessarily needs to be complex contrary to the case where $\delta$ is small. Without resonance, it was found in Sec.~\ref{sec:nonres} that $A_N$ needs to {be} several orders larger than $A_\lambda$ for correct amount of $\varepsilon\simeq\mathcal{O}(10^{-6})$. However, the resonance effect at $\delta=\pi$ mitigates this requirement allowing $A_N$ to be of the same order or smaller than $A_\lambda$.
\begin{figure}[h]
\begin{center}
\includegraphics[scale=0.5]{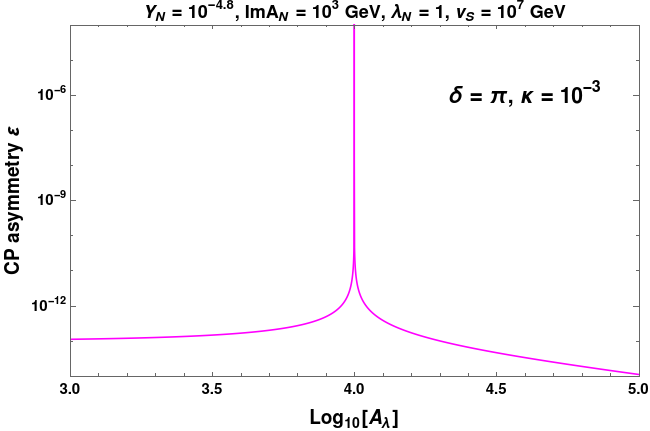}
\caption{Variation of $\varepsilon$ {versus} $A_\lambda$ in the $\delta = \pi$ limit. The resonance in $\varepsilon$ occurs when $A_\lambda = \kappa v_S$.}
\label{fig:res}
\end{center}
\end{figure}
\section{Conclusion}\label{sec:conclude}
We have presented a new mechanism for soft leptogenesis in the context of the NMSSM with a singlet {RH} neutrino superfield. Similar to soft leptogenesis in {the} MSSM, we also generate CP asymmetry at the tree{-}level owing to the CP violation occuring due to the difference between the mass and CP eigenstates similar to the $K^0-\bar{K}^0$ or the $B^0-\bar{B}^0$ systems. The difference lies in the fact that MSSM soft leptogenesis requires using thermal masses and phase space factors for boson and fermion final states without which there is no asymmetry. In the NMSSM where the singlet scalar $S$ takes a vev, an asymmetry can be generated even {without any thermal phase space/mass corrections to the decay products}. Further if there is spontaneous CP violation in the system with $\sin\delta\neq 0$, lepton asymmetry can be created without using any other complex parameter. In the numerical analysis for small $\delta$ case, we considered the mass scale of the RH sneutrino to be $10^7-10^8$~GeV. We found that to generate sufficient asymmetry, one of the soft trilinear coupling $A_N$ needs to be $\gtrsim 10^{7}$~GeV and $\kappa\lesssim \mathcal{O}(10^{-5})$. This also tells us that $A_\lambda\simeq \mathcal{O}(10^2)$~GeV. However, if $\delta\rightarrow \pi$, there occurs a resonance in the system which helps to obtain $\epsilon\sim \mathcal{O}(10^{-6})$ even with $A_N\lesssim A_\lambda$ and the value of $\kappa$ can be comparatively larger ($\gtrsim \mathcal{O}(10^{-3})$). The mass of the {RH} sneutrino came out in the range $\mathcal{O}(10^7-10^8)$~GeV which lies below the cosmological gravitino overproduction bound of $T_{{\rm reheat}}\simeq \mathcal{O}(10^9)$~{GeV}~\cite{Davidson:2002qv,Kawasaki:2006hm, Khlopov:1984pf, Falomkin1984}. This mass scale for the sneutrino (which depends on the vev of $S$, $v_S$) in {the} NMSSM also could favour gravitational wave detection at LIGO~\cite{Dev:2016feu} provided a strong first order phase transition occurs in the scalar sector. It would be interesting to explore the flavor effects in the present scenario that we leave for a future study. 

\vspace{0.5cm}
{ \noindent\Large \textbf{Acknowledgements}}

\vspace{0.2cm}
WA would like to acknowledge the support in the form of funding
available from the Department of Atomic Energy, Government of India for the
Neutrino Project at Harish-Chandra Research Institute (HRI). AK would like to thank Bharti Kindra for discussions on $K^0-\bar{K}^0$ systems. AK and AKS acknowledge PRL for  providing postdoctoral research fellowship.

\bibliographystyle{apsrev4-1}
\bibliography{refs}
\end{document}